\documentclass[aps,preprint,showpacs,superscriptaddress]{revtex4}%
\usepackage{amsfonts}
\usepackage{amsmath}
\usepackage{amssymb}
\usepackage{graphicx}%
\usepackage[dvipsnames]{xcolor}
\def\ag{\mbox{$\alpha$-$\gamma$} }

\begin{document}

\title{Momentum-dependent susceptibilities and magnetic exchange in bcc 
iron from supercell DMFT calculations}

\author{A. S. Belozerov}
\affiliation{M. N. Mikheev Institute of Metal Physics, Russian Academy of Sciences, 620137 Yekaterinburg, Russia}
\affiliation{Ural Federal University, 620002 Yekaterinburg, Russia}

\author{A. A. Katanin}
\affiliation{M. N. Mikheev Institute of Metal Physics, Russian Academy of Sciences, 620137 Yekaterinburg, Russia}
\affiliation{Ural Federal University, 620002 Yekaterinburg, Russia}

\author{V. I. Anisimov}
\affiliation{M. N. Mikheev Institute of Metal Physics, Russian Academy of Sciences, 620137 Yekaterinburg, Russia}
\affiliation{Ural Federal University, 620002 Yekaterinburg, Russia}


\begin{abstract}
We analyze 
the momentum- and temperature dependences of the magnetic susceptibilities and magnetic exchange interaction in paramagnetic bcc iron
by 
a combination of density functional theory and dynamical mean-field theory
(DFT+DMFT). 
By considering a general derivation of the orbital-resolved effective model for spin degrees of freedom
for Hund's metals,
we relate momentum-dependent susceptibilities
in the paramagnetic phase
to the magnetic exchange.
We then calculate 
non-uniform orbital-resolved susceptibilities
at high-symmetry wave vectors by constructing appropriate supercells in the DMFT approach. 
Extracting the irreducible parts of susceptibilities with respect to Hund's exchange interaction, we determine the corresponding orbital-resolved exchange interactions, which are then interpolated to the whole Brillouin zone. Using the spherical model we estimate
the temperature dependence of 
the resulting exchange between local moments.

\end{abstract}
\maketitle

\section{Introduction}

Iron has played an important role in the technological development of humankind and it is still an essential ingredient of modern industry. 
At low temperatures, iron is stabilized in the $\alpha$ phase with a body-centered cubic~(bcc) lattice.
When heated, iron first undergoes a magnetic transition to a paramagnetic state at a Curie temperature of 1043~K.
Further heating to 1185~K leads to a structural phase transition to the paramagnetic $\gamma$~phase with a face-centered cubic (fcc) lattice.
At 1660~K the latter again transforms to the bcc structure ($\delta$ phase), 
which is stable up to a melting point of 1810~K.

Despite extensive studies of iron, there is still a lack of detailed understanding of its electronic, magnetic, and structural properties.
Mainly, this is due to the difficulty in constructing a unified theory, which accurately takes into account Coulomb correlations and describes both, the itinerant and localized behavior of electrons.

The rapid development of computer technology in the last decades has made it feasible to study real materials using the so-called \emph{ab initio} techniques, the most successful of which are based on density functional theory (DFT). 
The DFT calculations of iron within the local density approximation (LDA) or generalized gradient approximations (GGA) resulted in a good description of the magnetic moments and energies of
the magnetically ordered ground state~\cite{Fe_DFT}.
%
%
The paramagnetic state of iron was simulated by a combination of DFT with a disordered local moment (DLM) model~\cite{dlm_method}, a spin-spiral approach~\cite{Ruban},
and a spin-space averaging procedure~\cite{Dusseldorf0}.
%
These approaches accurately captured the magnetic exchange interactions \cite{Lichtenstein84,Lichtenstein85_87,Antropov1999,Spisak1997,Frota_Pess2000,Pajda2001,KL,Shallcross2005,Gornostyrev,Kvashnin2015},
the Curie temperature~\cite{DLM_Tc}, and the phonon spectra~\cite{Ruban,Dusseldorf0}, and were used to study the energetics of the bcc-fcc lattice transformation~\cite{Okatov09,Zhang11}.
A combination of DFT with the Heisenberg model allowed one to consider correlation effects at finite temperatures~\cite{Gornostyrev1} and to describe the pressure dependence of the Curie temperature~\cite{{Dusseldorf2}}, the temperature dependence of magnon-phonon coupling~\cite{Dusseldorf5}, and thermodynamic properties~\cite{Dusseldorf134,Gornostyrev1}. 

An accurate treatment of electron correlations and local spin dynamics, which can be especially important at finite
temperatures, requires however
a combination of DFT with dynamical mean-field theory (DMFT)~\cite{DMFT}.
%
%
This combination is called the DFT+DMFT approach~\cite{LDA+DMFT} and 
allows one to carefully take into account local quantum fluctuations. 
%
%
%
The application of DFT+DMFT to iron resulted in accurate values of the magnetic moments both in the ferromagnetic and paramagnetic states~\cite{Lichtenstein2001,OurAlpha0,OurGamma,Igoshev,Chioncel2003,Grechnev2007,Kvashnin2015}.  
%
%
%
In $\alpha$-Fe the formation of local magnetic moments in the $e_g$ band, accompanied by the non-quasiparticle form of the electronic states in this band~\cite{OurAlpha0}, as well as the non-Fermi-liquid behavior of 
$t_{2g}$ states \cite{Igoshev} were obtained. It was shown that the local moments in bcc Fe are present up to the Earth core conditions \cite{Abrikosov,Abrikosov1}.
%
%
A substantial overestimation of the Curie temperature found in the first studies~\cite{Lichtenstein2001} was later shown to be due to violating the SU(2) symmetry of the Coulomb interaction~\cite{Belozerov2013,Belozerov_UJ,Sangiovanni} and neglecting non-local correlations~\cite{Katanin2016}. 
%
%
%
%
By means of the DFT+DMFT approach, important information about the structural properties and phase transitions of iron was obtained~\cite{Leonov_alpha_gamma,Leonov_phonons,Pourovskii2014,Glazyrin}.

The DFT+DMFT studies, however, account for the local correlations only, and apart from the local quantities, only a uniform susceptibility was extracted as a derivative of the magnetization over the external magnetic field~\cite{Lichtenstein2001,OurAlpha0,Belozerov2013,Igoshev,Abrikosov,Abrikosov1}.
This leaves the important question of the origin of magnetic exchange in iron, 
especially in the paramagnetic phase. By studying the momentum dependence of a bare bubble, the effective spin-fermion model allowed one to find the dominant contribution of $t_{2g}$-$e_g$ hybridized states to the magnetic exchange, yielding a nearest-neighbor Heisenberg exchange interaction in the paramagnetic state \cite{Igoshev}. This result was confirmed by more recent DFT and DFT+DMFT studies of the ordered phase \cite{Katsnelson}, in which the correlation effects are suppressed. 
In addition, an essential effect of non-local corrections, caused by the magnetic exchange, on the Curie temperature and energy of the $\alpha$  phase near the \ag transition was demonstrated within the DFT method combined with the Heisenberg model \cite{Gornostyrev1} and the DFT+DMFT approach combined with the spin-fermion model \cite{Katanin2016}.

The results of Ref. \cite{Igoshev}, where the magnetic exchange in the paramagnetic phase was analyzed, however, did not account for a possible vertex correction, as well as the orbital structure of Hund's exchange interaction. To study the (orbital-resolved) effects of the non-local vertex corrections, one can consider
non-uniform magnetization, which appears in the system as a reaction to the introduced non-uniform external magnetic field. 
Such a field can be introduced by extending the single-impurity model, considered in DMFT, to an appropriate supercell~\cite{Potthoff1999_1,Potthoff1999_2}. In contrast to the cellular DMFT ~\cite{clusters1,clusters2,clusters3}, this method does not introduce off-diagonal (with respect to the lattice sites) self-energy components, 
but it is much simpler computationally
and can be used to study the leading non-local correlation effects beyond DMFT.

In this paper, we calculate the non-uniform magnetic susceptibilities at high-symmetry wave vectors for paramagnetic bcc iron by the supercell DMFT approach, and use them to derive the effective orbital-dependent model for the spin degrees of freedom by extracting the  irreducible parts of susceptibilities with respect to Hund's exchange interaction (Secs.~\ref{SectIIA}, \ref{SectIIB}). This allows us to determine the orbital-resolved exchange interactions (Sec.~\ref{SectIIC}) and estimate the temperature dependence of the resulting exchange between local moments (Sec.~\ref{SectIII}). 
In Sec.~\ref{sec:conclusions} we present a summary and conclusions.

\section{Effective spin model and magnetic exchange}
\label{spin_model}

\subsection{The effective spin model and irreducible susceptibilities}
\label{SectIIA}
Before turning to the \textit{ab initio} approach, let us discuss the general derivation of the effective
model for spin degrees of freedom for systems, in which Hund's interaction plays a dominant role, coined Hund's metals. To this end we consider 
a general action, 
\begin{equation}
S[c,c^+]=S_{\rm el}[c,c^{+}]-\int_0^\beta d\tau \sum\limits_{i,mm^{\prime}}I^{mm^{\prime}}\mathbf{s}_{im}\mathbf{s}_{i m^{\prime}},
\end{equation}
where $c_{im\sigma},c^{+}_{im\sigma}$ are Grassmann variables, corresponding to the electronic degrees of freedom ($i$, $m$ and $\sigma$ are site-, orbital and spin indices), ${\bf s}_{im}=\sum_{\sigma \sigma'}c^+_{im\sigma} \mbox {\boldmath $\sigma $}_{\sigma \sigma'} c_{im \sigma'}/2$
are the electronic spin operators (\mbox {\boldmath $\sigma $} are the Pauli matrices), $I^{mm'}$ is the Hund's exchange interaction, $S_{\rm el}[c,c^{+}]$ includes the kinetic term and all electron interactions, apart from the Hund's one, and all variables are assumed to depend on the imaginary time  ${\tau\in[0,\beta=1/T]}$. The corresponding partition function is given by $Z=\int D[c,c^+] \exp(-S[c,c^+])$.
We decouple the Hund's interaction part of the Hamiltonian by the Hubbard-Stratonovich transformation:  
\begin{align}
H_\textrm{H}  &  =-\sum\limits_{i,mm^{\prime}}I^{mm^{\prime}}\mathbf{s}_{im}\mathbf{s}_{im^{\prime}}\nonumber\\
&  \rightarrow \sum\limits_{i,mm^{\prime}}
I^{mm'}\mathbf{S}_{im}\mathbf{S}_{im^{\prime}}%
-2\sum\limits_{i,mm^{\prime}}I^{mm'}\mathbf{S}_{im}\mathbf{s}_{im^{\prime}},%
\label{HH}
\end{align}
where ${\bf S}_{im}$ are the new spin variables, which are integrated over in the partition function. 
Integrating out the fermionic fields, in the static limit we arrive at the effective action, which up to quadratic order in $\mathbf{S}$
variables represents a 
classical Heisenberg-like model (cf. Ref. \cite{Katanin2016}), supplemented by the remaining local interaction term, originating from Hund's exchange interaction,
\begin{equation}
H=-\frac{1}{2}\sum_{i\neq j,mm^{\prime}}J_{ij}^{mm^{\prime}}\bar{\mathbf{S}}_{im}\bar{\mathbf{S}}_{jm^{\prime}}+\sum_{i,mm^{\prime}}(I^{mm^{\prime}}-\frac{1}{2} J_{ii}^{mm'})%
\bar{\mathbf{S}}_{im}\bar{\mathbf{S}}_{im^{\prime}},\label{H}%
\end{equation}
where $\bar{\bf S}_{im}=T\int_0^{\beta} {\bf S}_{im}d\tau $ is the static component of the field $\bf S$. 
Equation (\ref{H}) can be derived either by expanding the partition function in the fields ${\bf S}$ or expressing the propagator $X_{\bf q}^{mm'}=(\beta/N)
\sum_{ij} \langle \bar{S}^a_{im}
\bar{S}^a_{jm'}
\rangle e^{i {\bf q}({\bf R}_j-{\bf R}_i)}$ 
through the static non-uniform electronic susceptibilities  $\chi^{mm'}_{\mathbf q}=(1/N)\int_0^\beta d\tau \sum_{ij} \langle s^z_{im}(0) s^z_{jm'}(\tau)  \rangle e^{i {\bf q}({\bf R}_j-{\bf R}_i)}$ by
$X_{\bf q}^{mm'}=  
[2I^{mm'}]^{-1}+\chi^{mm'}_{\bf q}$, as follows straightforwardly from the Eq. (\ref{HH}). The static limit corresponds to weak imaginary time dependence of the electronic susceptibility, e.g., in the presence of local moments. We assume in the following that higher-order terms in the fields ${\bf S}$ depend essentially only on the local degrees of freedom ${\mathbf S}_{im} {\mathbf S}_{im'}$, similarly to the Hertz-Moriya-Millis theory \cite{MoriyaBook,RoschRev} and the unified spin fluctuation (USFL) approach \cite{MoriyaUn,MoriyaUn1,MoriyaBook}, such that the quadratic Hamiltonian (\ref{H}) describes correctly the non-local interactions. In our approach these higher order terms are considered also as static and are viewed as providing a soft constraint, which restricts the length of fields $\bar{\bf S}$ (cf. Refs. \cite{MoriyaUn,MoriyaUn1}).

The 
resulting orbital-resolved exchange interaction $J_{ij}^{mm'}$ has a Ruderman-Kittel-Kasuya-Yosida (RKKY)-like form, and its Fourier transform reads%
\begin{equation}
{J}_{\mathbf{q}}^{mm^{\prime}}=2I^{mm^{\prime\prime}}\left(
\chi_{\mathbf{q}}^{m^{\prime\prime}m^{\prime\prime\prime}}\right)
_{\mathrm{irr}}I^{m^{\prime\prime\prime}m^{\prime}},%
\label{Eq:Jq}
\end{equation}
where the summation (i.e. matrix product) over repeated indices is assumed and 
the (transverse) irreducible parts of non-uniform electronic susceptibilities $\chi^{mm'}_{\mathbf q}$
with respect to
Hund's exchange interaction 
are introduced:%
\begin{equation}
\left(  \chi_{\mathbf{q}}^{mm^{\prime}}\right)  _{\mathrm{irr}}=\left[
\left(  2\chi_{\mathbf{q}}^{mm^{\prime}}\right)  ^{-1}+I^{mm^{\prime}%
}\right]  ^{-1},%
\label{chi_irr}
\end{equation}
where $\left[  ...\right]  ^{-1}$ denotes the matrix inverse with respect to 
the orbital indices, and the factor of $2$ accounts for the difference between the
transverse and longitudinal susceptibilities.
\begin{figure}[t]
\includegraphics[clip=false,width=0.7\textwidth]{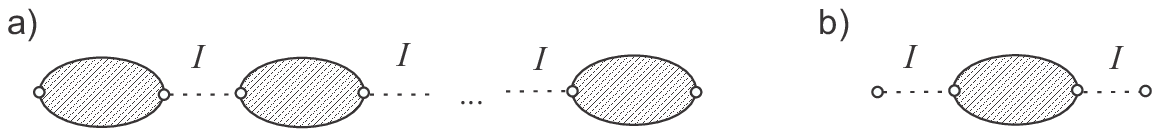}
\caption{
\label{fig:diag_susc}
Diagrammatic form of the non-uniform susceptibility $\chi_{\mathbf q}$ (a) and the result Eq. (\ref{Eq:Jq}) for the exchange interaction $J_{\mathbf q}$ (b). The bubbles represent the irreducible part $(\chi_{\mathbf q})_{\rm irr}$, and the dashed lines correspond to Hund's interaction $I$. The effects of the other types of interactions (as well as self-energy corrections) are absorbed in the irreducible susceptibility.}
\end{figure}
From the diagrammatic point of view, $(\chi_{\mathbf q})_{\rm irr}$ accounts for all vertex and self-energy corrections to the bubble, excluding Hund's interaction, connecting bubbles with each other (see Fig. \ref{fig:diag_susc}). We note that, apart from the static limit, Eqs. (\ref{H}) and (\ref{Eq:Jq}) represent formally the exact form of the quadratic in the new fields $\mathbf{S}$ terms of the Hamiltonian, which result from Hund's interaction, with the unknown quantities $(\chi_{\mathbf q}^{mm'})_{\rm irr}$, which are related to the electronic susceptibilities $\chi_{\mathbf q}^{mm'}$ via Eq. (\ref{chi_irr}). Similarly to ``unrenormalized" exchange interaction approaches, in our case the magnetic exchange is proportional to the irreducible susceptibility, and not its inverse \cite{Antropov2003} (cf. Ref. ~\cite{Bruno2003}). The advantage of the present approach in comparison with the USFL approach is that we consider Hund's interorbital interaction instead of the Hubbard intraorbital part, and the former does not yield ambiguities when using Hubbard-Stratonovich decoupling and accounts for the orbital structure of the interaction. 


\subsection{Supercell calculation of susceptibilities in DFT+DMFT }
\label{SectIIB}

%
The non-uniform susceptibility can be obtained by calculating a response to a small staggered external field introduced in the DMFT part in a suitable supercell,
namely, the orbital-resolved magnetic susceptibility $\overline{\chi}_{\mathbf{Q}_i}^{mm^{\prime}}
=4 \mu_B^2 {\chi}_{{\mathbf{Q}_i}}^{mm^{\prime}}=
\partial M_{\mathbf{Q}_i}^{m^{\prime}}/\partial H_{\mathbf{Q}_i}^m$, where 
$H_{\mathbf{Q}_i}^m$ is the magnetic field applied to the orbital~$m$ and
corresponding to the wave vector $\mathbf{Q}_i$, and
$M_{\mathbf{Q}_i}^{m^{\prime}}$ is the magnetization of orbital $m^{\prime}$.
In real space, the applied field
takes a form
${\mathbf{H}_{\mathbf{R}_j}^{m,i} = \mathbf{H}_0\, \textrm{cos}(\mathbf{Q}_i \mathbf{R}_j)}$,
where 
$\mathbf{R}_j$ is the position vector of site $j$,
and $\mathbf{H}_0$ is a constant small field.
In practice, we have used the magnetic field corresponding to splitting of the single-electron energies by 0.02 eV.  This field was checked to provide a linear response and was considered to be small enough to neglect the redistribution of charge density on the DFT level. 

For high-symmetry wave vectors, the corresponding supercells are compact,
and therefore can be studied by 
the real-space extension
of DMFT (see, e.g., Refs. ~\cite{Potthoff1999_1,Potthoff1999_2}).
In this extension, the self-energy is still local but it is assumed to be site-dependent.
As a result, several single-impurity problems have to be solved at each self-consistency loop. Note that neglect of the non-local components of the self-energies may yield an underestimation of the non-local components of the susceptibility. We expect, however, that because of strong on-site electronic correlations, non-local components of the self-energy do not change substantially the obtained results.


Let us turn to calculations for bcc iron.
%
First, 
we have performed DFT calculations using the full-potential linearized augmented-plane wave method implemented in the ELK code supplemented by the Wannier function projection procedure (Exciting-plus code).
%
The Perdew-Burke-Ernzerhof form of 
GGA was considered.
The calculations were carried out with the experimental lattice constant ${a=2.91}$~\AA\ in the vicinity of the \ag transition~\cite{Seki2005}.
The convergence threshold for total energy was set to $10^{-6}$~Ry.
The integration in reciprocal space was performed using an 18$\times$18$\times$18\, $\textbf{k}$-point mesh for the unit cell, while\, 15$\times$15$\times$15\,, and 12$\times$12$\times$12\, meshes were used for supercells with 2 and 4 atoms, respectively.
From converged DFT results we have constructed effective Hamiltonians in the basis of Wannier functions, which were built as a projection of the original Kohn-Sham states to site-centered localized functions as described in Ref.~\onlinecite{Korotin08}, considering $3d$, $4s$ and $4p$ states.

In DMFT calculations we use the Hubbard parameter ${U\equiv F^0=4}$~eV
and Hund's rule coupling ${I\equiv (F^2+F^4)/14=0.9}$~eV,
where $F^0$, $F^2$, and $F^4$ are the Slater integrals as obtained in
Ref.~\onlinecite{Belozerov_UJ} by the constrained DFT in the basis of $spd$ Wannier functions.
The on-site Coulomb interaction was considered in the density-density form.
The corresponding matrix of Hund's exchange can be expressed via the Coulomb interaction matrix as $I^{mm'}=(U^{mm'}_{\sigma,-\sigma}%
-U^{mm'}_{\sigma,\sigma})(1-\delta_{mm'})$. 
The double-counting correction was taken in the fully localized limit.
The impurity problem was solved by the hybridization expansion continuous-time quantum Monte Carlo method~\cite{CT-QMC}.

\begin{figure}
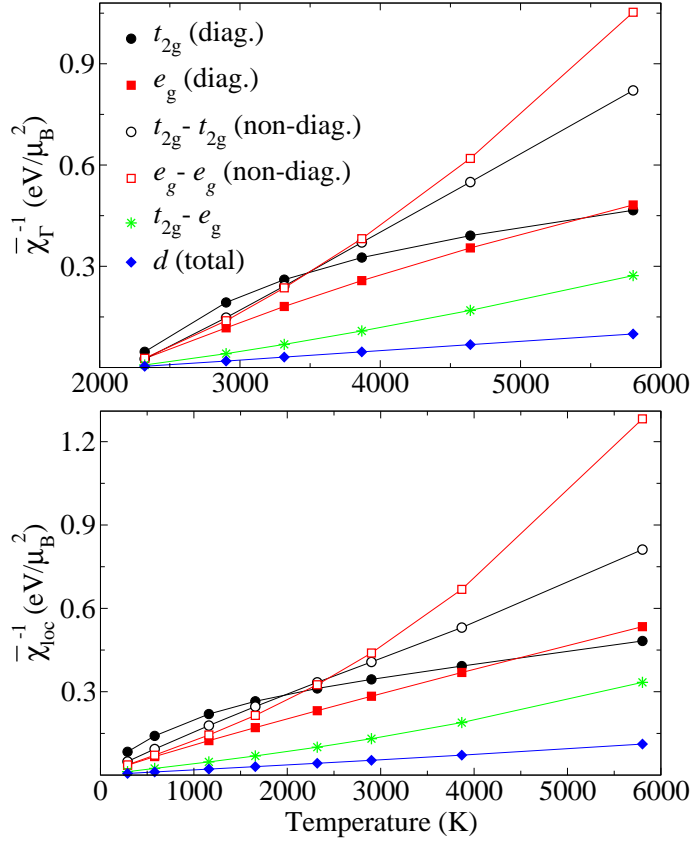

\includegraphics[clip=true,width=0.55\textwidth]{Fig2a_susc_gamma_uniform_diag_nondiag.eps}
\\
\vspace{0.15cm}
\includegraphics[clip=true,width=0.55\textwidth]{Fig2b_susc_loc_2.eps}
\caption{(Color online)
Temperature dependence of inverse uniform $\overline{\chi}_{\Gamma=\mathbf{0}}^{mm^{\prime}}$
(top panel) and local $\overline{\chi}^{mm'}_{\rm loc}$
(bottom panel) magnetic susceptibilities obtained by DFT+DMFT.
The diagonal and non-diagonal orbital-resolved contributions, as well as the total susceptibility for $d$ states are presented. 
\label{fig:loc_and_uniform_susc}}
\end{figure}

%
As a first step, let us consider the orbital-resolved uniform magnetic susceptibility,  as well as local magnetic susceptibility 
$\overline{\chi}^{mm'}_{\rm loc}=4\mu_B^2 {\chi}_{\rm loc}^{mm'}$, where ${\chi}_{\rm loc}^{mm'}= \int\nolimits_{0}^{\beta}\langle{s^z_{im}}(\tau){s^z_{im'}}(0)\rangle d\tau$.
%
The temperature dependence of inverse susceptibilities is 
presented in Fig.~\ref{fig:loc_and_uniform_susc}.
In both cases, only diagonal $t_{2g}$ and non-diagonal \mbox{$e_g$-$e_g$} contributions
show a significant non-linear dependence on temperature.
%
For the former contribution, this is in agreement with a previously found deviation from the Fermi-liquid behavior of $t_{2g}$ states~\cite{Igoshev}.
The non-linear behavior of the former (latter) contribution corresponds to an increase (decrease) of effective local moments with temperature, while the inverse total susceptibility is almost linear. 
This indicates that the above mentioned contributions compensate each other and may be closely related. 
The Curie temperature, obtained from the extrapolation of the inverse uniform susceptibility is 
${T_\textrm{C}\approx 2150}$~K (${\beta\approx 5.4}$~eV$^{-1}$).

To calculate the non-uniform susceptibilities we have constructed supercells containing up to 4 atoms and corresponding to high-symmetry points $\textrm{H}$, $\textrm{N}$, and $\textrm{P}$.
In particular, for wave vectors ${\mathbf{Q}_\textrm{H}=\{2\pi,0,0\}/a}$
and ${\mathbf{Q}_\textrm{N}=\{\pi,\pi,0\}/a}$,
we considered
supercells containing two nearest-neighbor atoms
at $(0,0,0)$ and $(a/2,a/2,a/2)$ in Cartesian coordinates.
%
For the supercell corresponding to ${\mathbf{Q}_\textrm{H}}$, the lattice vectors are
${\{a,0,0\}}$,
${\{0,a,0\}}$,
${\{0,0,a\}}$, while 
for ${\mathbf{Q}_\textrm{N}}$, they are
${\{-a,a,a\}/2}$,
${\{a,-a,a\}/2}$,
${\{a,a,0\}}$.
%
For ${\mathbf{Q}_\textrm{P}=\{\pi,\pi,\pi\}/a}$,
we built a supercell with 4 atoms by including 2 extra atoms
at $(a,0,0)$ and $(-a/2,a/2,a/2)$.
The lattice vectors for this supercell are 
${\{-a,0,a\}}$,
${\{2a,0,0\}}$,
${\{-a,a,0\}}$.
%

\subsection{Results for irreducible susceptibilities and orbital-resolved magnetic exchange}
\label{SectIIC}

\begin{figure}[t]
\includegraphics[clip=false,width=0.54\textwidth]
{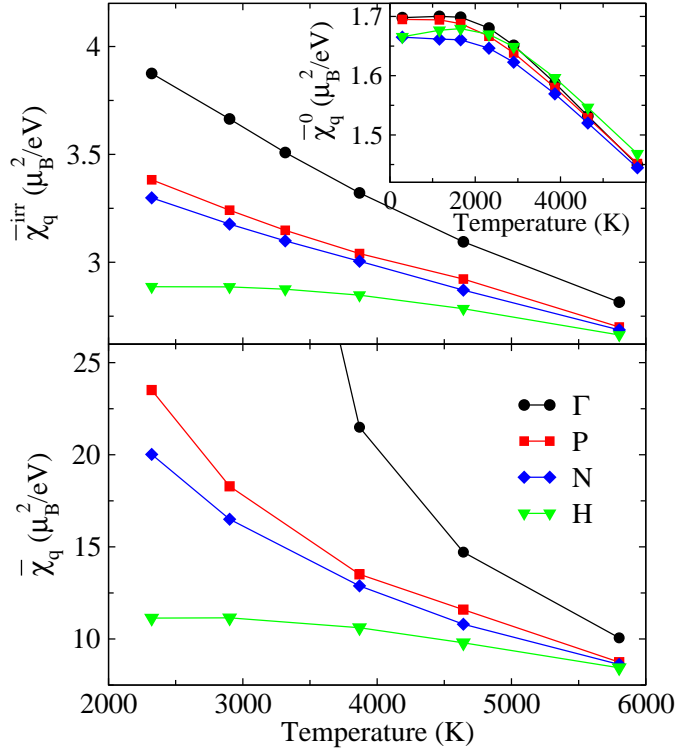}
\caption{(Color online)
\label{fig:model_susc1}
Temperature dependence of the orbital-summed irreducible magnetic susceptibility ${\overline{\chi}_\textbf{q}^{\rm irr}=2\mu_B^2 \sum_{m,m'} ({\chi}_\textbf{q}^{mm'})_{\rm irr}}$ (top panel), particle-hole bubble
$\overline{\chi}_\textbf{q}^{0}$
(inset), and the susceptibility in the supercell approach (bottom panel) within DMFT.
}
\end{figure}

\begin{figure}[t]
\includegraphics[clip=false,width=0.54\textwidth]{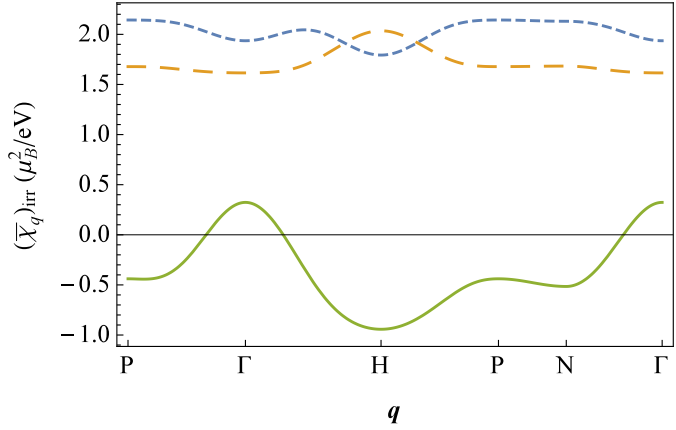}
\caption{(Color online)
\label{fig:susc_q}
Momentum dependence of the irreducible magnetic susceptibility $\overline{\chi}_\textbf{q}^{\rm irr}$
at $\beta=5$ eV$^{-1}$ for 
$e_g$ states 
(blue short-dashed line), $t_{2g}$ states (yellow long-dashed line), and the hybridized $t_{2g}$-$e_g$
states (green solid line). 
}
\end{figure}

The temperature dependence of the orbital-summed irreducible susceptibilities at ${T>T_\textrm{C}}$ is shown in Fig.~\ref{fig:model_susc1}, which is compared to the calculation of the bubble 
$\overline{\chi}_\textbf{q}^{0} = -(2\mu_B^2/\beta) \sum_{\textbf{k},n,m,m'} G^{mm'}_\textbf{k} (i\omega_n) G^{m'm}_\textbf{k+q}(i\omega_n)$ and full momentum-dependent susceptibility in the supercell approach. One can see that the major part of the divergence of the uniform susceptibility $\overline{\chi}_\Gamma$ and strong enhancement of the other non-uniform susceptibilities near the magnetic phase transition are removed when passing to the irreducible susceptibilities, yet the obtained irreducible susceptibilities are approximately twice larger than the corresponding values of the bubble, 
which can be attributed to vertex corrections and the effects of the other components of the Coulomb interaction, apart from the Hund's term. 
The irreducible susceptibility $(\chi_{\mathbf q})_{\rm irr}$ also shows a somewhat stronger, although qualitatively similar momentum dependence, compared to that of the bubble at low temperatures (cf. Ref. \cite{Igoshev} and the inset in Fig.~\ref{fig:model_susc1}).
In contrast to the bubble, the maximum of the obtained irreducible susceptibility is at the $\Gamma$ point for all considered temperatures, which favors ferromagnetic correlations.
The momentum dependence of the irreducible susceptibility 
${\overline{\chi}_\textbf{q}^{\rm irr}=2\mu_B^2 \sum_{m\in M,m'\in M'}({\chi}_\textbf{q}^{mm'})_{\rm irr}}$, summed over groups of orbitals, corresponding to $e_g$ ($M=M'=e_g$) states, $t_{2g}$ ($M=M'=t_{2g}$) states, and the hybridized $t_{2g}$-$e_g$ states ($M=t_{2g}$ ($e_g$), $M'=e_g$ ($t_{2g}$)) at $\beta=5$ eV$^{-1}$ is shown in Fig.~\ref{fig:susc_q}, where
the interpolation scheme between different points ${\bf Q}_i$, outlined below is used (cf. Ref. \cite{Igoshev} for the momentum dependence of the bubble at low $T$).

While the components of the exchange interaction ${J}_{\mathbf{Q}_{i}}$ can be determined from the obtained irreducible susceptibilities, to
interpolate between different points $\mathbf{Q}_{i}$  we consider an expansion 
\begin{align}
{J}_{\mathbf{q}}^{mm^{\prime}} &  ={J}^{mm^{\prime},(0)}+\overline{J}_{\mathbf{q}%
}^{mm^{\prime}},\\
\overline{J}_{\mathbf{q}}^{mm^{\prime}} &  =J^{mm^{\prime},(1)}\cos(aq_{x}/2)\cos
(aq_{y}/2)\cos(aq_{z}/2)\nonumber\\
&  +J^{mm^{\prime},(2)}\left[\cos(aq_{x})+\cos(aq_{y})
+\cos(aq_{z})\right]\nonumber\\
&  +J^{mm^{\prime},(3)}\left[\cos(aq_{x})\cos(aq_{y})+\cos(aq_{y})\cos(aq_{z})
+\cos(aq_{z})\cos(aq_{x})\right],
\end{align}
where $J^{mm^{\prime},(r)}$ are determined from $J_{\mathbf{Q}_{i}%
}^{mm^{\prime}}.$ Since there are four high-symmetry points 
$(\Gamma,\textrm{H},\textrm{P},\textrm{N})$ for the bcc lattice,
they are sufficient to determine exchange
interactions up to third next-nearest neighbors $(r=3)$. More explicitly,%
\begin{align}
J^{(0)} &  =\frac{1}{8}(J_{\text{$\Gamma$}}+J_{\text{H}}+6J_{\text{N}%
}),\nonumber\\
J^{(1)} &  =\frac{1}{2}(J_{\text{$\Gamma$}}-J_{\text{H}}),\nonumber\\
J^{(2)} &  =\frac{1}{12}(J_{\text{$\Gamma$}}+J_{\text{H}}-2J_{\text{P}%
}),\nonumber\\
J^{(3)} &  =\frac{1}{24}(J_{\text{$\Gamma$}}+J_{\text{H}}+4J_{\text{P}%
}-6J_{\text{N}}).
\end{align}


\begin{figure}[t]
\includegraphics[width=0.55\textwidth]{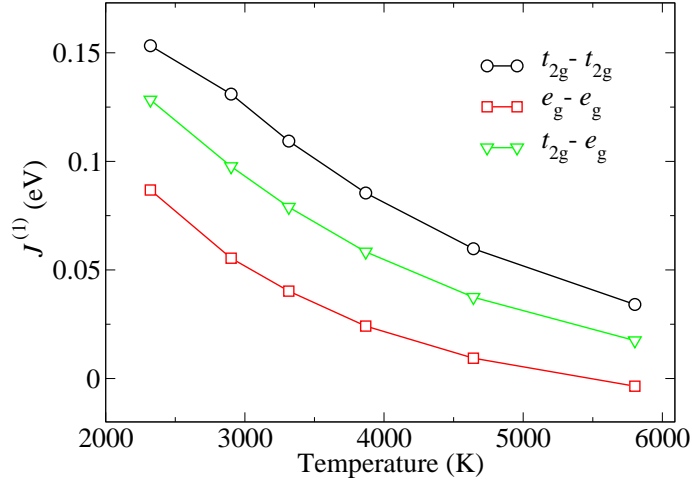}
\caption{(Color online)
\label{fig:exch1}
Temperature dependence of the exchange parameter $J^{(1)}$, representing the nearest-neighbor exchange integral between the electronic local moment states of different symmetries, multiplied by the number of nearest neighbors.
}
\end{figure}

The site-diagonal part of the interaction, $J{^{(0)}\sim 1}$~eV, is found to be weakly temperature dependent. The temperature dependence of the most important exchange interaction $J^{(1)}$, which characterizes the nearest-neighbor exchange interaction, multiplied by the number of nearest neighbors $z_1=8$,
is presented in Fig. \ref{fig:exch1}.
The interactions $J^{(2,3)}$ are much smaller and are not shown here, although they were accounted for in the actual calculations.
The largest exchange is obtained between $t_{2g}$ states, a somewhat smaller one between $t_{2g}$ and $e_{g}$ states, and the smallest exchange is between $e_{g}$ states. Note that the participation of $t_{2g}$ states (apart from more localized $e_g$ states) in the local moment formation was suggested earlier in Ref. \cite{Igoshev}. 
One should, however, distinguish the symmetry of the {\it local moment} states, which provide the largest contribution to the exchange, discussed here, and the symmetry of the {\it itinerant} states, which mediate this contribution; for the latter $t_{2g}$-$e_g$ hybridized states were suggested to be dominant \cite{Igoshev,Katsnelson}. 
In our approach these states also yield the most dispersive contribution to the susceptibility (see Fig. \ref{fig:susc_q}), and therefore are expected to provide
the largest itinerant contribution.





\section{Spherical approximation to magnetic exchange}
\label{SectIII}

To obtain the 
physically observable exchange interaction between local moments,
we resolve the soft constraint on the length of fields $\bar{\mathbf{S}}$, discussed in Sec.~\ref{SectIIA}, 
by using, similarly to Ref.  \cite{Katanin2016}, 
 the spherical model approximation, which fixes respective moments in different orbitals equal to their DMFT values, $\langle\bar{\mathbf{S}}_{im}\bar{\mathbf{S}}_{im'}\rangle_{H}=3[d(1/\chi_{\rm loc}^{mm'})/dT]^{-1}.
$
To this end we consider the corresponding action of the orbital-dependent spherical model, obtained by a combination of the Hamiltonian (\ref{H}) with the contributions originating from the above mentioned constraints and proportional to the new fields $i \lambda^{mm'}$, 
\begin{equation}
\mathcal{S}_{\rm H}=\frac{1}{2}\sum\limits_{\mathbf{q},mm'}\int d\tau\left[
(i \lambda^{mm'}-
\overline{J}_{\mathbf{q}}^{mm^{\prime}})\bar{\mathbf{S}}_{\mathbf{q},m}%
\bar{\mathbf{S}}_{-\mathbf{q},m^{\prime}}
-i\lambda^{mm'}
\langle\bar{\mathbf{S}}_{\mathbf{q},m}\bar{\mathbf{S}}_{-\mathbf{q},m'}\rangle \right].  \label{SSph}%
\end{equation}
where $\bar{\mathbf{S}}_{\mathbf{q},m}$ is the Fourier transform of $\bar{\mathbf{S}}_{im}$ and we absorb all momentum-independent terms into $\lambda^{mm'}$. We consider the saddle point approximation $i\lambda^{mm'}=\lambda^{mm'}_{0}$ (such that the corresponding $\lambda$-dependent terms can be also viewed as appearing from the decoupling of fourth-order and higher terms in $\mathbf{S}$-fields, 
cf. Refs. \cite{MoriyaUn,MoriyaUn1,Katanin2016}).
Integrating over the fields $\bar{\mathbf S}_{\mathbf{q},m}$, 
we obtain
\begin{align}
\ln Z &  =\ln\int D[{\bf S}]\exp(-\mathcal{S}_{\rm H})\nonumber\\
&  =-\frac{3}{2}\sum\limits_{\mathbf{q}}\ln\det\left(  \lambda_{0}^{mm'}%
-\overline{J}_{\mathbf{q}}^{mm^{\prime}}\right)
+\frac{1}{2T}\sum_{mm'}\lambda_{0}^{mm'}\langle\bar{\mathbf{S}}_{im}\bar{\mathbf{S}}_{im'}\rangle.
\end{align}
From this we find the respective averages%
\begin{equation}
\langle\bar{\mathbf{S}}_{im}\bar{\mathbf{S}}_{im'}\rangle_{H}=3T\sum\limits_{\mathbf{q}}\left[
\lambda_{0}^{mm'}
-\overline{J}_{\mathbf{q}}^{mm^{\prime}%
}\right]^{-1}  
\end{equation}
which yield
non-linear equations for 
the parameters $\lambda_{0}^{mm'}%
$. The introduced quantities $\lambda_0^{mm'}$ play the role of Moriya $\lambda$-correction~\cite{MoriyaUn1,MoriyaBook}, which was previously used in a similar fashion in the USFL theory \cite{MoriyaUn,MoriyaUn1} and more recently in D$\Gamma$A approach \cite{Katanin2009}. These quantities describe the (orbital-dependent) ``renormalization" 
of propagators of the spin fields ${\bf S}$, 
such that these propagators fulfill the sum rules, required for local parts of the susceptibilities. Physically this implies ascribing to the fields $\mathbf{S}$, introduced in section \ref{SectIIA}, the meaning of the local spin moment.

To determine the effective exchange interaction we 
introduce the total on-site spin $\mathbf{S}_{\mathbf{q}}%
=\sum\nolimits_{m}\bar{\mathbf{S}}_{\mathbf{q}m}$ and require the equivalence of the
spherical model considered here to that of Ref. \cite{Katanin2016},%
\begin{align}
\langle\mathbf{S}_{\mathbf{q}}\mathbf{S}_{-\mathbf{q}}\rangle_{H} &
=3T\sum\limits_{mm^{\prime}}\left[  \lambda_{0}^{mm'}
-\overline{J}_{\mathbf{q}}^{mm^{\prime}}\right]  ^{-1}  =\frac{3T}{\lambda_{0}-J_{\mathbf{q}}},%
\end{align}
where $\lambda_{0}$ is some constant, which can be determined from the
condition of the absence of self-interaction, $\sum\nolimits_{\mathbf{q}%
}J_{\mathbf{q}}=0.$

\begin{figure}[h]
\includegraphics[width=0.6\textwidth]{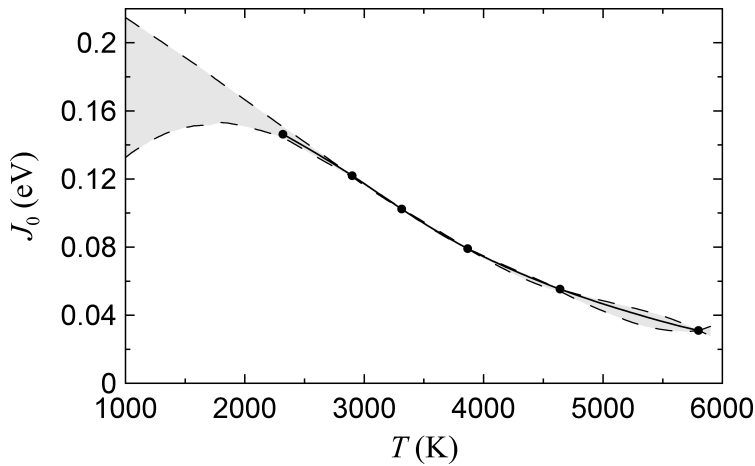}
\caption{
\label{fig:exch}
Exchange interaction $J_0$ at different temperatures (points). The solid line shows the interpolation, and the shaded area shows the possible range of interpolations and extrapolation to low temperatures.}
\end{figure}

The temperature evolution of the resulting exchange interaction is shown in Fig.~\ref{fig:exch}. Because of a correct account of the matrix structure of Hund's interactions and susceptibilities, including vertex corrections, which increase the values of irreducible susceptibilities, the resulting low-temperature exchange interaction
$J_0
\simeq 0.20$~eV
is somewhat larger than the value of $0.13$~eV, obtained previously in the paramagnetic phase \cite{Igoshev} for the same value of the Coulomb interaction, but  remains comparable to $0.18$~eV, previously obtained for $U=2.3$eV in Ref. \cite{Igoshev}.
Comparing the obtained temperature dependence of $J_0$ to the results of Fig. \ref{fig:exch1}, one can see, that the $t_{2g}$-$t_{2g}$ exchange gives the major contribution to $J_0$. 

To compare the obtained value of $J_0$ to the DFT results for $\alpha$-iron \cite{Lichtenstein85_87,Antropov1999,Spisak1997,Frota_Pess2000,Pajda2001,KL,Kvashnin2015,Shallcross2005,Gornostyrev}, one has to multiply it by half of the square of the spin $\mu_{\rm eff}^2/(8\mu_B^2)$, since the DFT exchange interactions are attributed to unit classical spin vectors [the extra factor 1/2 comes from our definition of the Hamiltonian in Eq. (\ref{H})].
For the magnetic moment extracted from the uniform susceptibility in our DMFT calculation, $\mu_{\rm eff}^2=9.2\mu_B^2$ (which is close to the experimental value $\mu_{\rm eff}^2=9.8\mu_B^2$),
we get
a somewhat larger value of the magnetic exchange than that obtained in the ferromagnetic state and spin spiral states
\cite{Lichtenstein85_87,Antropov1999,Spisak1997,Frota_Pess2000,Pajda2001,KL,Kvashnin2015,Shallcross2005,Gornostyrev,Kvashnin2015,Katsnelson}, but in reasonable agreement with the renormalized exchange interaction in the ferromagnetic state \cite{KL} and the result of the DLM approach \cite{Shallcross2005}.

The obtained exchange interaction can also be verified against the known results for the spin stiffness constant of $\alpha$-iron. For crystals with cubic symmetry, the spin wave spectrum near the $\Gamma$ point behaves as $\omega_\textbf{q}=D\textbf{q}^2$, where for the dominant nearest-neighbor exchange interaction $D=J_0\, a^2 \mu_\textrm{eff}/(2z_1\mu_\textrm{B})$. The experimental values of $D$, obtained by magnetization~\cite{Pauthenet1982} and neutron scattering~\cite{Shirane1968,Stringfellow1968,Mook1973} measurements, range from 280 to 330~meV$\cdot\textup{\AA}^2$ at low temperatures, which corresponds to $J_0$ 
from 0.17 to 0.21 eV, in agreement with our estimates.

\section{Conclusion}
\label{sec:conclusions}

We have proposed the derivation of the effective orbital-resolved Heisenberg-like model of magnetic exchange, which originates from Hund's interaction, assuming the latter provides the dominant contribution to the magnetic exchange. 
We have determined the parameters of this model for paramagnetic bcc iron by calculating the non-uniform magnetic susceptibilities in appropriate supercells within DMFT combined with DFT.
We have found that the momentum dispersion and temperature dependence of the irreducible (with respect to Hund's exchange interaction) susceptibility is similar to that of the bare particle-hole bubble at low temperatures, although somewhat enhanced due to correlation effects; in contrast to the bubble, the obtained irreducible susceptibility has always a maximum at the $\Gamma$ point, favoring ferromagnetic correlations. 

From the obtained irreducible susceptibilities we have extracted individual components of magnetic exchange interaction, showing that the exchange between $t_{2g}$-$t_{2g}$ states, and $t_{2g}$-$e_{g}$ states provide the largest contribution.
To extract the resulting (physical) exchange interaction, we have considered the spherical model approach, similar to the one studied previously for iron in Ref.  \cite{Katanin2016}. The obtained value of the exchange interaction at low-temperatures $J_0
\sim
0.20$ eV is close to previous estimates in the renormalized magnetic force ~\cite{KL} and DLM \cite{Shallcross2005} approaches, and slowly decreases with temperature, dropping twice at $T\sim 3000$~K.

The proposed approach can be used in other substances with local  moments and a dominant role of Hund's exchange interaction (so-called Hund's metals), including a possibility of studying magnetoelastic coupling. At the same time, the extension of the presented approach to the weak itinerant (nearly) ferro- and antiferromagnets, in particular, studying the properties of the $\gamma$ phase of iron, which is known to be more itinerant than the $\alpha$ phase, especially at low temperatures \cite{OurGamma}, is of certain interest.

The proposed approach can be also further used to study non-local corrections beyond DMFT in multi-orbital systems, since it does not require evaluation of the local vertices for obtaining the non-uniform susceptibility. In this respect, 
the evaluation of the corresponding non-local contributions to the electronic self-energy on the basis of the obtained static or analogously calculated dynamic susceptibilities has to be explored.

\begin{acknowledgments}
The work was supported by the Russian Science Foundation (Project No. 14-22-00004).
\end{acknowledgments}


\begin{thebibliography}{9}                                                           
\bibitem{Fe_DFT} 
  D. J. Singh, W. E. Pickett, and H. Krakauer, Phys. Rev. B \textbf{43}, 11628 (1991); 
  L. Stixrude, R. E. Cohen, and D. J. Singh, \textit{ibid.} \textbf{50}, 6442  (1994);
  E. G. Moroni, G. Kresse, J. Hafner, and J. Furthmuller, \textit{ibid.} \textbf{56}, 15629 (1997). 

\bibitem{dlm_method} B. L. Gyorffy, A. J. Pindor, J. Staunton, G. M. Stocks, and H. Winter, J. Phys. F: Met. Phys. \textbf{15}, 1337 (1985). 

\bibitem{Ruban} A. V. Ruban and V. I. Razumovskiy, Phys. Rev. B \textbf{85}, 174407 (2012).

\bibitem{Dusseldorf0} F. K\"ormann, A. Dick, B. Grabowski, T. Hickel, and
J. Neugebauer, Phys. Rev. B \textbf{85}, 125104 (2012).

  \bibitem{Lichtenstein84}
A. I. Liechtenstein, M. I. Katsnelson, and V. A. Gubanov, J. Phys. F: Met. Phys. \textbf{14}, L125 (1984).

\bibitem{Lichtenstein85_87}
A. I. Liechtenstein, M. I. Katsnelson, and V. A. Gubanov, Solid State Comm. \textbf{54}, 327 (1985);
A. I. Liechtenstein, M. I. Katsnelson, V. P. Antropov, and V. A. Gubanov, J. Magn. Magn. Mater. \textbf{67}, 65 (1987).

\bibitem{Antropov1999}
V. P. Antropov, B. N. Harmon, and A. N. Smirnov, J. Magn. Magn. Mater. \textbf{200}, 148 (1999);
M. van Schilfgaarde and V. P. Antropov, J. Appl. Phys. \textbf{85}, 4827 (1999).

\bibitem{Spisak1997}
D. Spi\v{s}\'{a}k and J. Hafner, J. Magn. Magn. Mater. \textbf{168}, 257 (1997).

\bibitem{Frota_Pess2000}
S. Frota-Pess\^oa, R. B. Muniz, and J. Kudrnovsk\'y, Phys. Rev. B \textbf{62}, 5293 (2000).

\bibitem{Pajda2001} M. Pajda, J. Kudrnovsk\'y, I. Turek, V. Drchal, and P. Bruno, Phys. Rev. B \textbf{64}, 174402 (2001).

\bibitem{KL} M. I. Katsnelson and A. I. Lichtenstein, J. Phys.: Condens. Matter {\bf 16}, 7439 (2004).

\bibitem{Shallcross2005}
S. Shallcross, A. E. Kissavos, V. Meded, and A. V. Ruban, Phys. Rev. B \textbf{72}, 104437 (2005).

\bibitem{Gornostyrev} S. V. Okatov, Yu. N. Gornostyrev, A. I. Lichtenstein, and
M. I. Katsnelson, Phys. Rev. B {\bf 84}, 214422 (2011).

\bibitem{Kvashnin2015} Y. O. Kvashnin, O. Gr\aa{}n\"{a}s, I. Di Marco, M. I. Katsnelson, A. I. Lichtenstein, and O. Eriksson, Phys. Rev. B \textbf{91}, 125133 (2015).


\bibitem{DLM_Tc} J. B. Staunton and B. L. Gyorffy, Phys. Rev. Lett. \textbf{69}, 371 (1992).

\bibitem{Okatov09} S. V. Okatov, A. R. Kuznetsov, Yu. N. Gornostyrev, V. N. Urtsev, and M. I. Katsnelson, Phys. Rev. B \textbf{79}, 094111 (2009).

\bibitem{Zhang11} H. Zhang, B. Johansson, and L. Vitos, Phys. Rev. B \textbf{84}, 140411(R) (2011).

\bibitem{Gornostyrev1} I. K. Razumov, D. V. Boukhvalov, M. V. Petrik, V. N. Urtsev, A. V. Shmakov, M. I. Katsnelson, and Yu. N. Gornostyrev, Phys. Rev. B {\bf 90}, 094101 (2014).

\bibitem{Dusseldorf2} F. K\"ormann, A. Dick, T. Hickel, and J. Neugebauer, Phys. Rev. B {\bf 79}, 184406 (2009).
  
\bibitem{Dusseldorf5} F. K\"ormann, B. Grabowski, B. Dutta, T. Hickel, L. Mauger, B. Fultz, and J. Neugebauer, Phys. Rev. Lett. {\bf 113}, 165503 (2014).  
  
\bibitem{Dusseldorf134}
  F. K\"ormann, A. Dick, B. Grabowski, B. Hallstedt, T. Hickel, and J. Neugebauer, Phys. Rev. B {\bf 78}, 033102 (2008);
  F. K\"ormann, A. Dick, T. Hickel, and J. Neugebauer, \textit{ibid.} {\bf 81}, 134425 (2010);
  T. Hickel, B. Grabowski, F. K\"ormann, and J. Neugebauer,  J. Phys.: Condens. Matter {\bf 24}, 053202 (2012). 

\bibitem{DMFT}
  W. Metzner and D. Vollhardt, Phys. Rev. Lett. \textbf{62}, 324 (1989);
  A. Georges, G. Kotliar, W. Krauth and M. J. Rozenberg, Rev. Mod. Phys. \textbf{68}, 13 (1996).

\bibitem{LDA+DMFT} 
  V. I. Anisimov, A. I. Poteryaev, M. A. Korotin, A. O. Anokhin, and G. Kotliar, J. Phys.: Condens. Matter \textbf{9}, 7359 (1997); 
  A. I. Lichtenstein and M. I. Katsnelson, Phys. Rev. B \textbf{57}, 6884 (1998).

\bibitem{OurGamma}P. A. Igoshev, A. V. Efremov, A. I. Poteryaev, A. A. Katanin, V. I. Anisimov, Phys. Rev. B \textbf{88}, 155120 (2013).

\bibitem{Chioncel2003} L. Chioncel, L. Vitos, I. A. Abrikosov, J. Koll\'ar, M. I. Katsnelson, and A. I. Lichtenstein, Phys. Rev. B \textbf{67}, 235106 (2003).

\bibitem{Grechnev2007}  A. Grechnev, I. Di Marco, M. I. Katsnelson, A. I. Lichtenstein, J. Wills, and O. Eriksson, Phys. Rev. B \textbf{76}, 035107 (2007).

\bibitem{Lichtenstein2001}
  A. I. Lichtenstein, M. I. Katsnelson, and G. Kotliar, Phys. Rev. Lett. \textbf{87}, 067205 (2001).  

\bibitem{OurAlpha0} A. A. Katanin, A. I. Poteryaev, A. V. Efremov, A. O. Shorikov, S. L. Skornyakov, M. A. Korotin, V. I. Anisimov, Phys. Rev. B \textbf{81}, 045117 (2010).

\bibitem{Igoshev}P. A. Igoshev, A. V. Efremov, A. A. Katanin, Phys. Rev. B \textbf{91}, 195123 (2015).

\bibitem{Abrikosov} L. V. Pourovskii, T. Miyake, S. I. Simak, A. V. Ruban, L. Dubrovinsky, and I. A. Abrikosov, Phys. Rev. B {\bf 87}, 115130 (2013).

\bibitem{Abrikosov1} O. Yu. Vekilova, L. V. Pourovskii, I. A. Abrikosov, and S. I. Simak, Phys. Rev. B {\bf 91}, 245116 (2015).

\bibitem{Belozerov2013} A. S. Belozerov, I. Leonov, and V. I. Anisimov, Phys. Rev. B \textbf{87}, 125138 (2013).

\bibitem{Belozerov_UJ} A. S. Belozerov and V. I. Anisimov,  J. Phys.: Condens. Matter \textbf{26}, 375601 (2014).

\bibitem{Sangiovanni} A. Hausoel, M. Karolak, E. Sasioglu, A. Lichtenstein, K. Held, A. Katanin, A. Toschi,
and G. Sangiovanni, Nature Comm. {\bf 8}, 16062 (2017).

\bibitem{Katanin2016}A. A. Katanin, A. S. Belozerov, and V. I. Anisimov, Phys. Rev. B \textbf{94}, 161117(R) (2016).

\bibitem{Leonov_alpha_gamma}
  I. Leonov, A. I. Poteryaev, V. I. Anisimov, and D. Vollhardt, Phys. Rev. Lett. \textbf{106}, 106405 (2011).

\bibitem{Leonov_phonons}
  I. Leonov, A. I. Poteryaev, V. I. Anisimov, and D. Vollhardt, Phys. Rev. B \textbf{85}, 020401(R) (2012);
  I. Leonov, A. I. Poteryaev, Yu. N. Gornostyrev, M. I. Katsnelson, V. I. Anisimov, and D. Vollhardt, Scientific Rep. \textbf{4}, 5585 (2014).

\bibitem{Pourovskii2014} L. V. Pourovskii, J. Mravlje, M. Ferrero, O. Parcollet, and I. A. Abrikosov, Phys. Rev. B \textbf{90}, 155120 (2014).

\bibitem{Glazyrin} K. Glazyrin, L. V. Pourovskii, L. Dubrovinsky, O. Narygina, C. McCammon, B. Hewener, V. Schünemann, J. Wolny, K. Muffler, A. I. Chumakov \textit{et al.}, Phys. Rev. Lett. \textbf{110}, 117206 (2013).

\bibitem{Katsnelson} Y. O. Kvashnin, R. Cardias, A. Szilva, I. Di Marco, M. I. Katsnelson, A. I. Lichtenstein, L. Nordstr\"om, A. B. Klautau, and O. Eriksson, Phys. Rev. Lett. {\bf 116}, 217202 (2016).

  
\bibitem{Potthoff1999_1}
M. Potthoff and W. Nolting, Phys. Rev. B \textbf{59}, 2549 (1999).

\bibitem{Potthoff1999_2}
M. Potthoff and W. Nolting, Phys. Rev. B \textbf{60}, 7834 (1999). 

\bibitem{clusters1} 
  A. I. Lichtenstein and M. I. Katsnelson, Phys. Rev. B \textbf{62}, R9283 (2000).
  

\bibitem{clusters2}
  G. Kotliar, S. Y. Savrasov, G. Palsson, and G. Biroli, Phys. Rev. Lett. \textbf{87}, 186401 (2001).
  
\bibitem{clusters3} T. Maier, M. Jarrell, T. Pruschke, and M. H. Hettler,
Rev. Mod. Phys. {\bf 77}, 1027 (2005).  


\bibitem{MoriyaBook} T. Moriya, ``Spin Fluctuations in Itinerant Electron Magnetism", Springer (Berlin), 1985.

\bibitem{RoschRev} H. v. L\"ohneysen, A. Rosch, M. Vojta, and P. W\"olfle, Rev. Mod. Phys. {\bf 79}, 1015 (2007).

\bibitem{MoriyaUn} T. Moriya and Y. Takahashi, J. Phys. Soc. Jpn. {\bf 45}, 397 (1978).

\bibitem{MoriyaUn1} T. Moriya, J. Magn. Magn. Mater. {\bf 14}, 1 (1979).

\bibitem{Antropov2003} V. P. Antropov, J. Magn. Magn. Mater. \textbf{262}, L192 (2003).

\bibitem{Bruno2003} P. Bruno, Phys. Rev. Lett. \textbf{90}, 087205 (2003).

\bibitem{Seki2005} I. Seki, K. Nagata, ISIJ Int. \textbf{45}, 1789 (2005).

\bibitem{Korotin08} Dm. Korotin, A. V. Kozhevnikov, S. L. Skornyakov, I. Leonov, N. Binggeli, V. I. Anisimov, and G. Trimarchi,  Eur. Phys. J. B \textbf{65}, 91  (2008).

\bibitem{CT-QMC} A. N. Rubtsov, V. V. Savkin, and A. I. Lichtenstein, Phys. Rev. B \textbf{72}, 035122 (2005);
P. Werner, A. Comanac, L. de Medici, M. Troyer, and A. J. Millis, Phys. Rev. Lett. \textbf{97}, 076405 (2006).
  
\bibitem{Katanin2009} A. A. Katanin, A. Toschi, and K. Held, Phys. Rev. B {\bf 80}, 075104 (2009).




\bibitem{Pauthenet1982} R. Pauthenet, J. Appl. Phys. \textbf{53}, 8187 (1982).

\bibitem{Shirane1968} G. Shirane, V. J. Minkiewicz, and R. Nathans, J. Appl. Phys. \textbf{39}, 383 (1968).

\bibitem{Stringfellow1968}
M. W. Stringfellow, J. Phys. C \textbf{1} 950 (1968).

\bibitem{Mook1973} H. A. Mook and R. M. Nicklow, Phys. Rev. B \textbf{7}, 336 (1973).


\end{thebibliography}
\end{document}